\documentclass[a4paper,11pt]{article}
\usepackage{pos}

\title{Recent results on hard and rare probes from ALICE}

\author*[a, b]{Alena Gromada}
\author{for the ALICE collaboration}

\affiliation[a]{GSI Helmholtzzentrum für Schwerionenforschung,\\
  Planckstra\ss e 1, Darmstadt, Germany}

\affiliation[b]{Physikalisches Institut, Ruprecht-Karls-Universit\"{a}t Heidelberg,\\
Im Neuenheimer Feld 226, Heidelberg, Germany}

\emailAdd{A.Gromada@gsi.de}
\manuallySeparateAuthors

\abstract{
	In high-energy hadronic collisions, hard parton scatterings with large momentum transfers are prerequisites for the formation of hard and rare probes. In heavy-ion collisions, these probes---final state particles related to the early hard-parton scatterings---serve as a powerful tool to explore the whole evolution of the medium including the quark--gluon plasma (QGP) stage. Hard probes in proton-proton (pp) collisions test perturbative quantum chromodynamics (pQCD) processes and hadronization, and provide a reference for the nuclear collision systems. High-multiplicity pp collisions provide a bridge to heavy-ion collisions, due to their large event activity.
	
	Heavy-flavor hadrons containing at least one charm or beauty quark belong to the hard and rare probes. Recently, ALICE has measured a broad palette of heavy-flavour baryons allowing to shed more light on charm fragmentation. Measurements of charmonium production as a function of the event multiplicity can provide insight into the interplay between charmonium-production processes and soft processes driving the multiplicity. ALICE has investigated the multiplicity dependence of J/$\psi$ production at midrapidity and the ratio of $\psi \rm (2S)$ to J/$\psi$ yields at forward rapidity. The charm and beauty cross sections can be constrained by fits of Monte-Carlo generators to the measured dielectron continuum.
	}

\FullConference{
  The Eighth Annual Conference on Large Hadron Collider Physics - LHCP2020 \\
  25-30 May, 2020\\
  online}

\begin{document}
\maketitle

\section{Introduction}
The study of heavy-flavor production in pp collisions has been crucial for testing pQCD. In this contribution, recent ALICE results on charm and beauty production in pp collisions are going to be discussed with a focus on those measured in collisions at the highest center-of-mass energy ever reached in the laboratory ($\sqrt{s} = 13$~TeV). The dataset collected by ALICE at this energy allows to study charm in high multiplicity environments with improved precision.

According to the factorization theorem, the heavy-flavor hadron production cross section can be obtained as a convolution of the parton distribution functions, the hard-scattering heavy-flavor $\rm q\bar{q}$ cross section at the partonic level, and the fragmentation function. The creation of the $\rm q\bar{q}$ pairs can be treated perturbatively down to zero transverse momentum, $p_{\rm T}$, due to the large masses of the heavy-flavor quarks, whereas the hadronization is considered to be, at least to a large extent, a soft process. Quarkonia hadronize through an evolution of a heavy-flavor $\rm q\bar{q}$ pair into a bound state. Open-heavy flavor hadrons are produced via the fragmentation of single heavy quarks into hadrons or via coalescence. Observables especially sensitive to the fragmentation processes are the relative yields of various species of charm baryons and mesons. Studies of heavy-flavor production as a function of the charged-particle multiplicity density could improve our understanding of processes at the partonic level.

\begin{figure}[h]
	\begin{minipage}{\textwidth}
		\centering
		\raisebox{-0.052\height}{\includegraphics[width=0.379\textwidth]{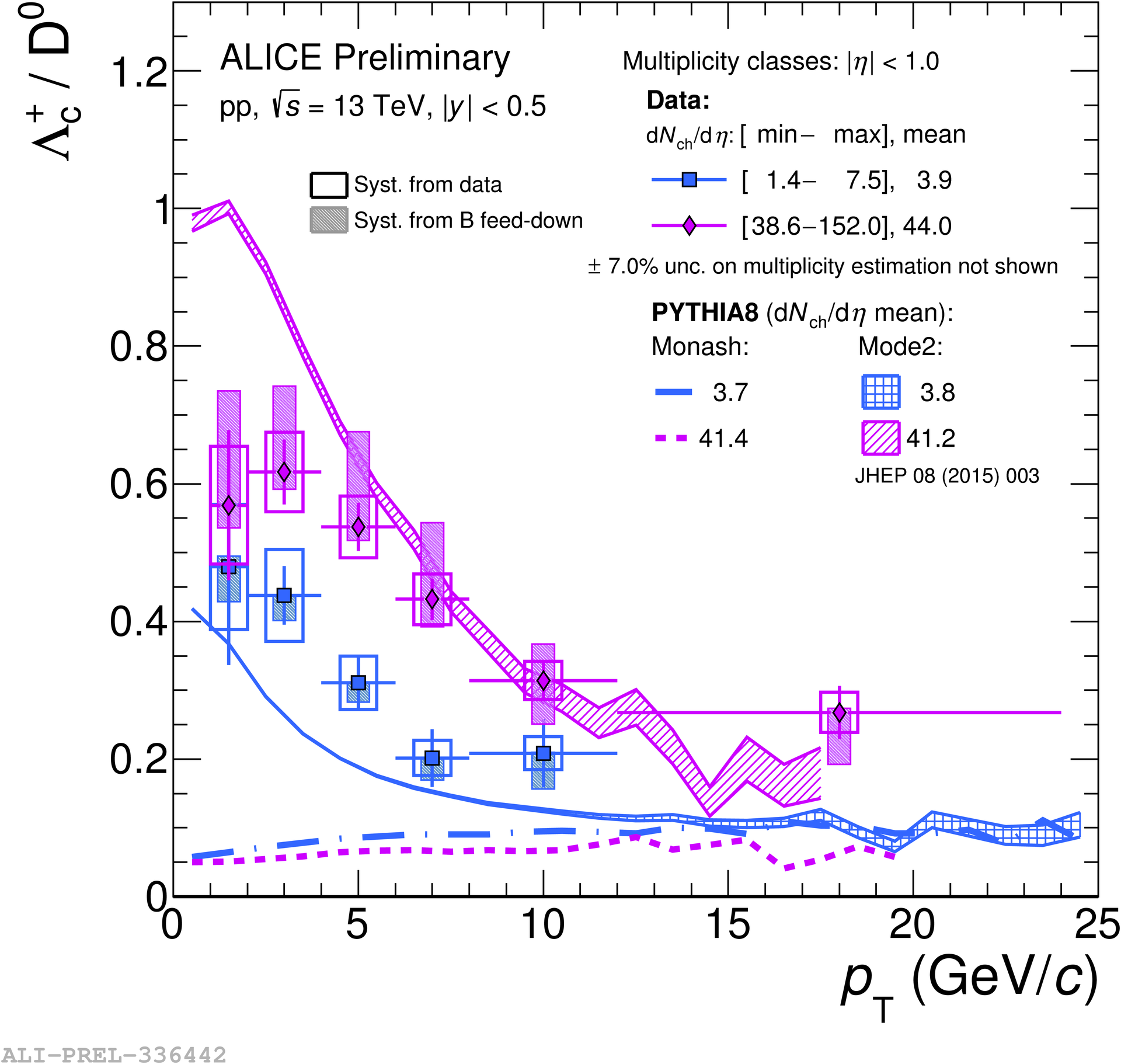}}
		\hspace{0.5cm}
		\raisebox{-0.0\height}{\includegraphics[width=0.433\textwidth]{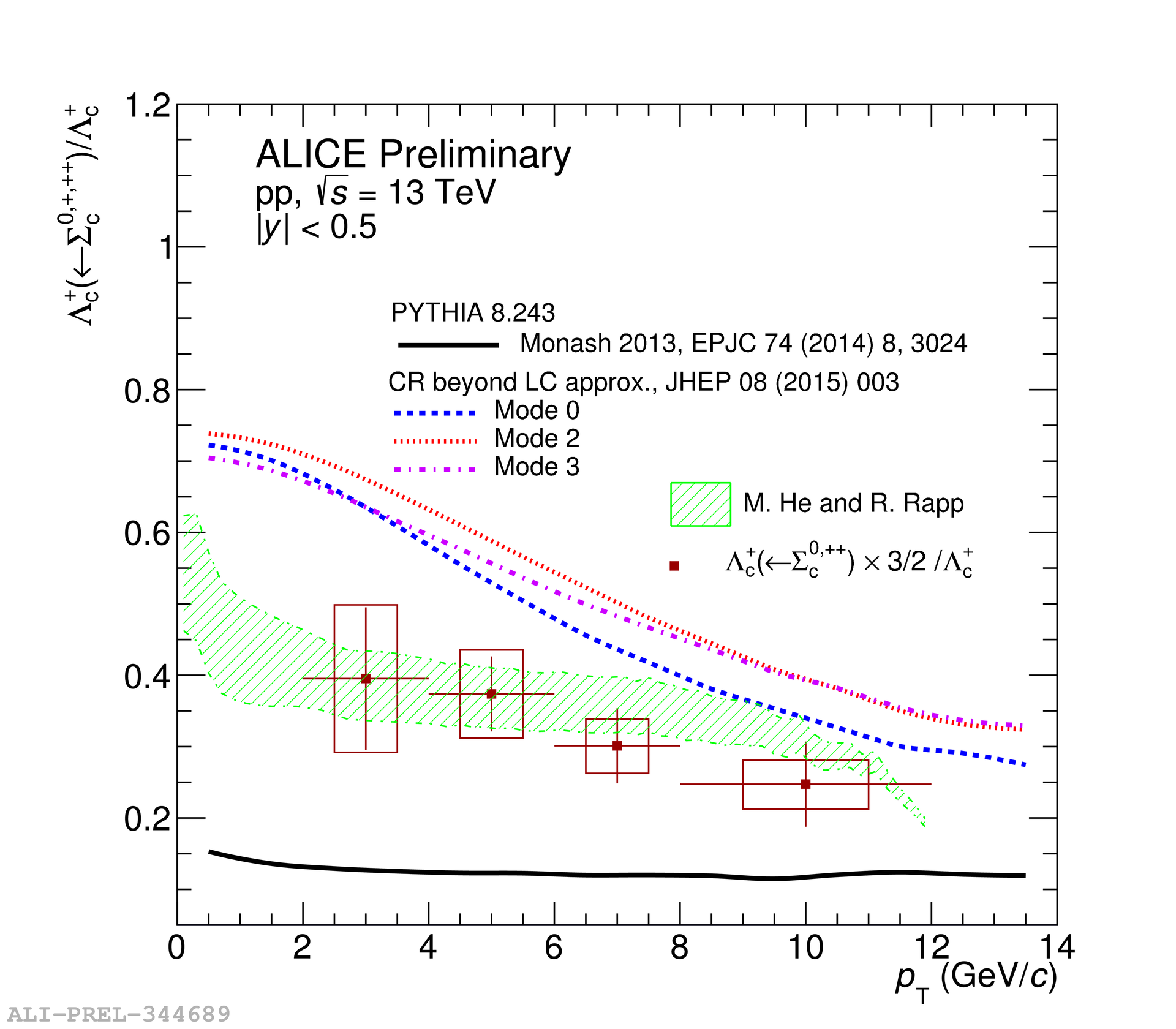}}
	\end{minipage}
	\caption{(Left) The $\Lambda_{\rm c}^{+}/\rm D^0$ ratio for two multiplicity classes and (right) the $\Lambda_{\rm c}^{+}(\leftarrow \Sigma_{\rm c})/\Lambda_{\rm c}^{+}$ ratio both measured in pp collisions at $\sqrt{s} = 13$~TeV, as a function of $p_{\rm T}$. }
	\label{fig:Baryons1}
\end{figure}

\section{Open-charm production}
With the dataset of pp collisions collected at the highest LHC center-of-mass energy $\sqrt{s} = 13$~TeV, ALICE measured the $\Lambda_{\rm c}^{+}/\rm D^0$ ratio for different multiplicity classes. Recently, it was extended to the highest-multiplicity class with the average $\rm{d}$$N_{\rm{ch}}$$/\rm{d}\eta$ corresponding to 44.0, well above the average minimum-bias pp multiplicity of 6.89. In the left panel of Fig.~\ref{fig:Baryons1}, the $\Lambda_{\rm c}^{+}/\rm D^0$ ratio for two $\rm{d} $$N_{\rm{ch}}$$/\rm{d}\eta$-classes is shown as a function of $p_{\rm T}$. For both multiplicity classes, the value of the ratio ranges between 0.4 and 0.6 at low $p_{\rm T}$ (1--4~GeV/$c$) and decreases with increasing $p_{\rm T}$ down to 0.2--0.3 in the probed $p_{\rm T}$ intervals. At low $p_{\rm T}$, the data is largely underestimated by PYTHIA Monash \cite{Skands_2014} with the charm fragmentation tuned on $\rm{e^+e^-}$ data. $p_{\rm T}$-trend of the data is described by PYTHIA8 with color reconnection (CR) beyond leading color (LC) \cite{Christiansen:2015yqa}, which allows for the rearrangement of string connections such that partons from different initial parton scattering processes can also interact. The model-data comparison hints that CR could play an important role in the charm fragmentation.

\begin{figure}[h]
	\begin{minipage}{\textwidth}
		\centering
		\raisebox{-0.0\height}{\includegraphics[width=0.411\textwidth]{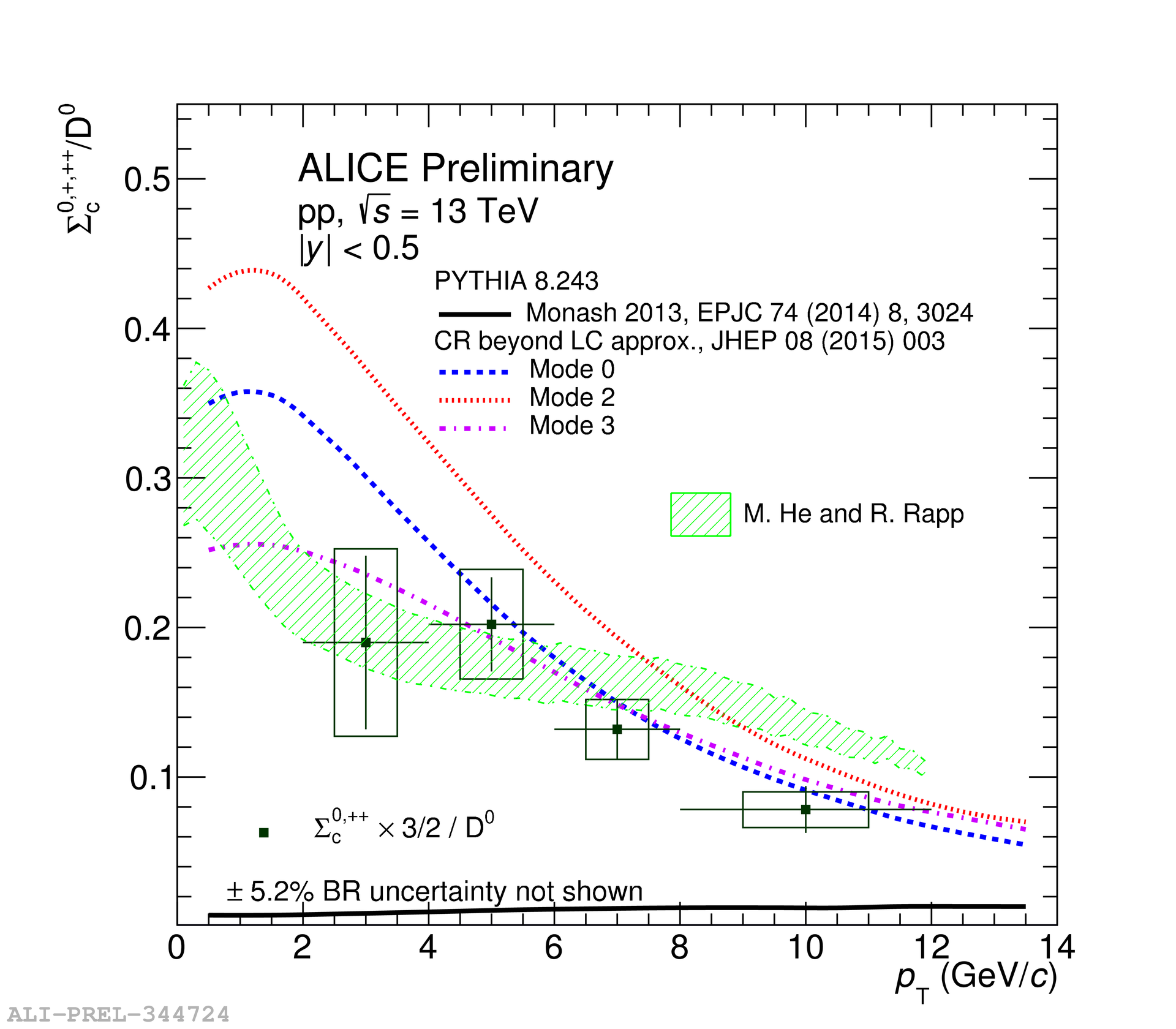}}
				\hspace{0.1cm}
		\raisebox{-0.034\height}{\includegraphics[width=0.386\textwidth]{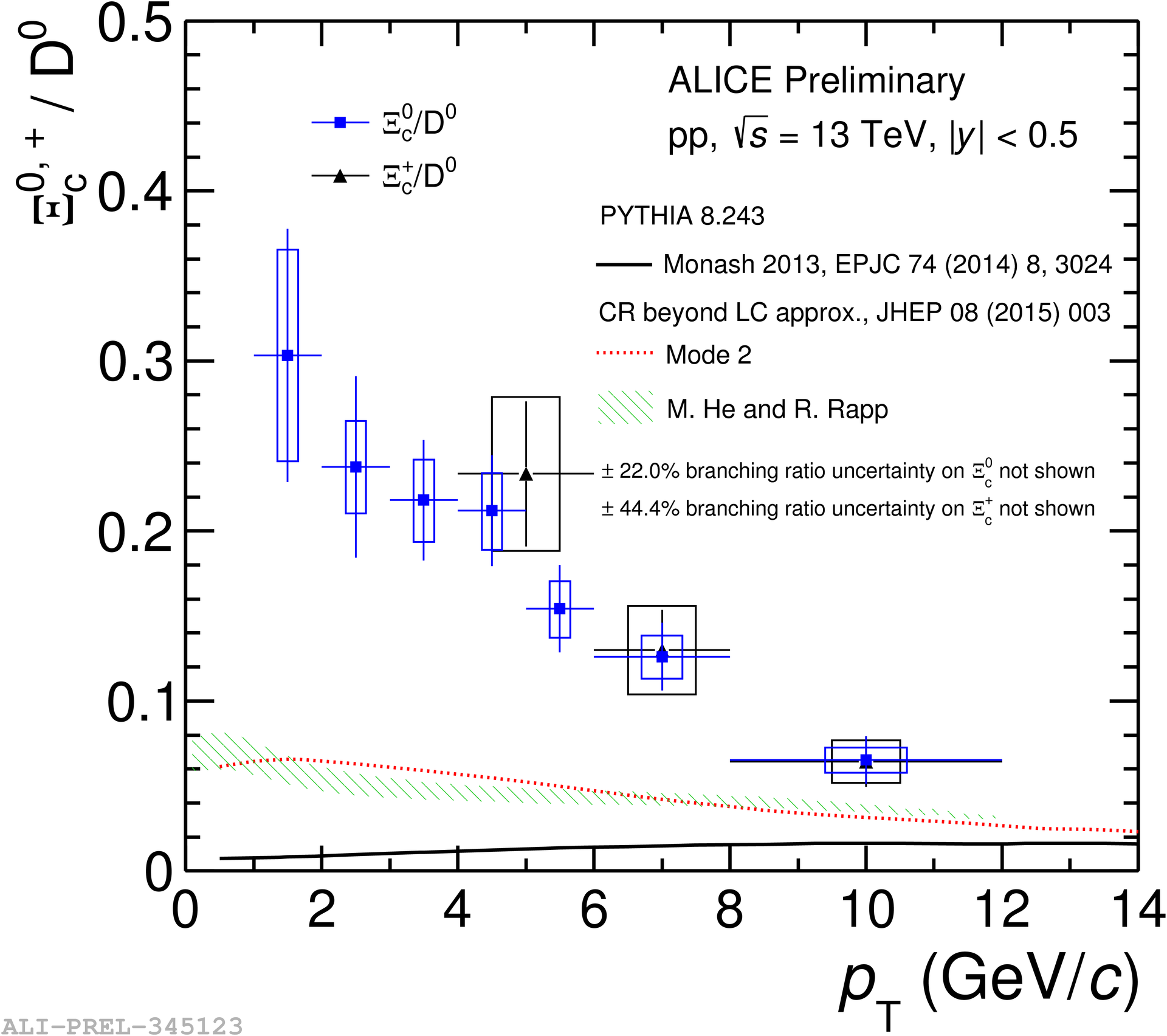}}
	\end{minipage}
	\caption{(Left) The $\Sigma_{\rm c}/\rm D^0$ ratio and (right) $\Xi_{\rm c}^{0}/\rm D^0$ and $\Xi_{\rm c}^{+}/\rm D^0$ ratios at $\sqrt{s} = 13$~TeV as a function of $p_{\rm T}$. }
	\label{fig:Baryons2}
\end{figure}

However, there are more mechanisms at play that could affect the $\Lambda_{\rm c}^{+}/\rm D^0$ ratio such as the feed-down from $\Sigma_{\rm c}$ states. ALICE has for the first time measured the $\Sigma_{\rm c}^{0}$ and $\Sigma_{\rm c}^{++}$ baryons by pairing soft pion candidates to reconstructed displaced $\Lambda_{\rm c}^{+}$ decay vertices.
As can be seen in the right panel of Fig.~\ref{fig:Baryons1}, showing the $\Lambda_{\rm c}^{+}(\leftarrow \Sigma_{\rm c})/\Lambda_{\rm c}^{+}$ ratio vs. $p_{\rm T}$, approximately 20 to 60\% of $\Lambda_{\rm c}^{+}$ baryons in the probed $p_{\rm T}$-intervals originate from $\Sigma_{\rm c}$ feed-down. The $\Sigma_{\rm c}/\rm D^0$ ratio shown in the left panel of Fig.~\ref{fig:Baryons2} hints at a similar $p_{\rm T}$-dependence as the $\Lambda_{\rm c}^{+}/\rm D^0$ ratio. Therefore, the $\Sigma_{\rm c}$ feed-down contribution might explain part of the $\Lambda_{\rm c}^{+}$ enhancement with respect to $\rm D^0$ production. A more precise measurement is needed to draw a stronger conclusion. Both ratios $\Lambda_{\rm c}^{+}(\leftarrow \Sigma_{\rm c})/\Lambda_{\rm c}^{+}$ and $\Sigma_{\rm c}/\rm D^0$ are underestimated by PYTHIA tuned on $\rm{e^+e^-}$ measurements, whereas different modes of CR beyond LC overestimate $\Lambda_{\rm c}^{+}(\leftarrow \Sigma_{\rm c})/\Lambda_{\rm c}^{+}$ and describe $\Sigma_{\rm c}/\rm D^0$. The ratios are in agreement with statistical hadronization model (SHM) calculations with an augmented set of charm-baryon states \cite{HE2019117}. PYTHIA underestimates also the $\Xi_{\rm c}^0/\rm D^0$ and $\Xi_{\rm c}^+/\rm D^0$ ratios, shown in the right panel of Fig.~\ref{fig:Baryons2}, supporting breaking of the fragmentation-function universality into baryons in leptonic and hadronic colliding systems. The ratios are largely underestimated as well by PYTHIA with CR beyond LC and SHM. This points to the fact that the $\Xi_{\rm c}$ production is underestimated and underlines the importance of a re-evaluation of the total charm cross section.

\begin{figure}[h]
	\begin{minipage}{\textwidth}
		\centering
		\raisebox{-0.0\height}{\includegraphics[width=0.37\textwidth]{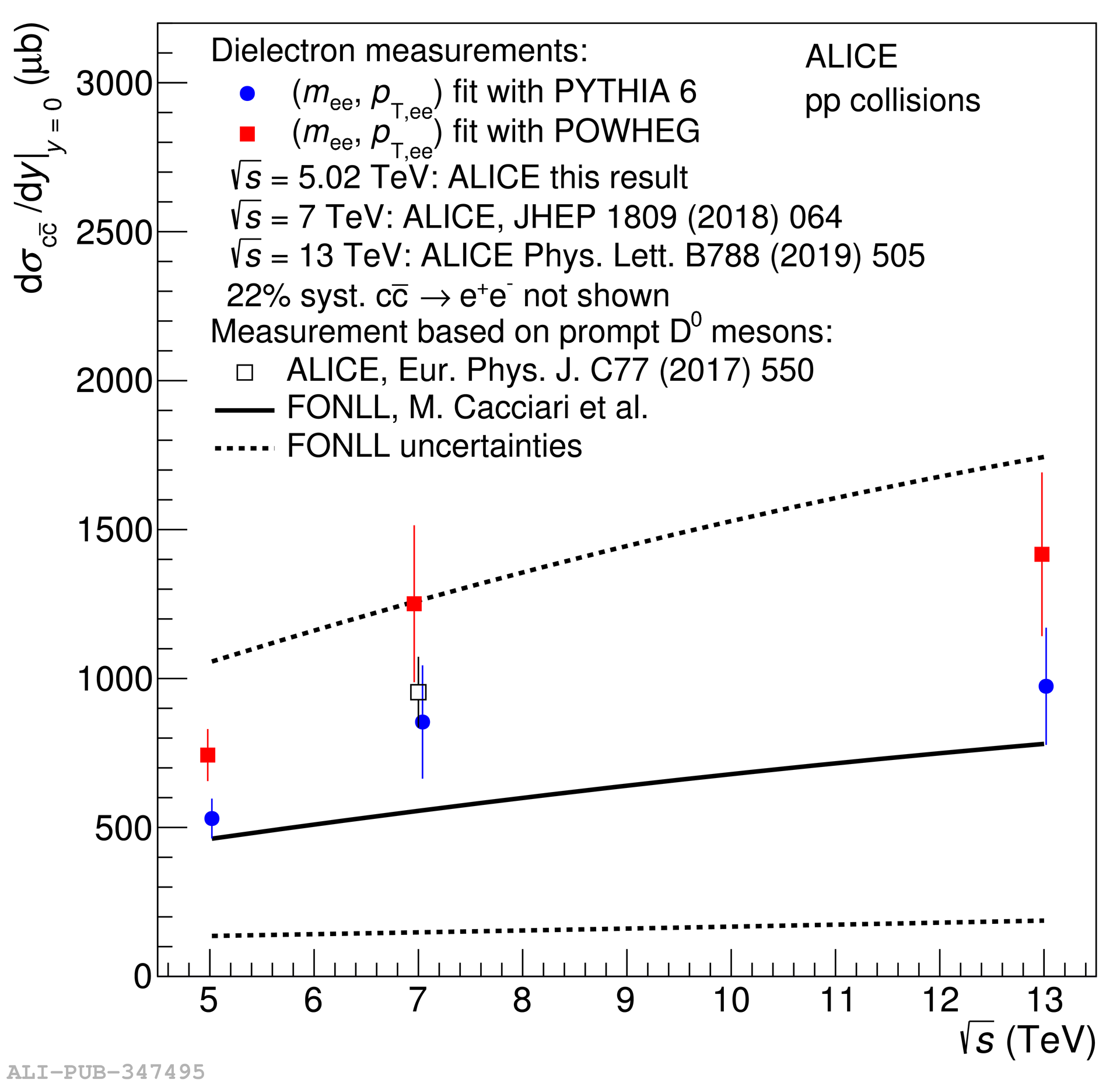}}
		\hspace{0.3cm}
		\raisebox{-0.0\height}{\includegraphics[width=0.37\textwidth]{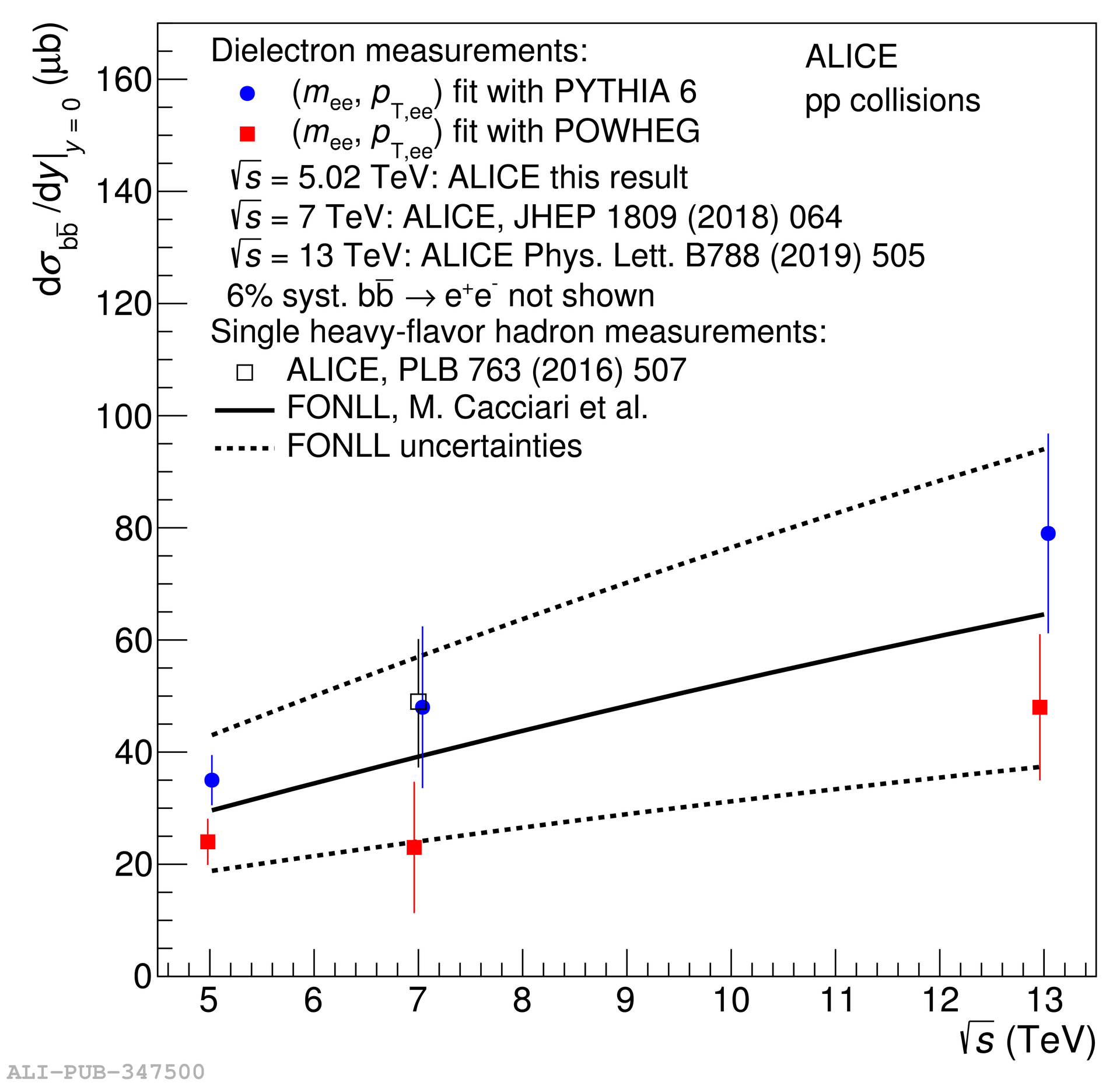}}
	\end{minipage}
	\caption{The total charm (left) and beauty (right) cross sections as a function of $\sqrt{s_{\rm}}$ in pp collisions. }
	\label{fig:Dielectrons}
\end{figure}

\section{Heavy-flavor cross section from the measured dielectron continuum}

The charm and beauty cross sections can be constrained by fits of Monte Carlo generators to the dielectron continuum. Figure~\ref{fig:Dielectrons} shows recently published cross sections for pp collisions at $\sqrt{s} = $ 5~TeV \cite{collaboration2020dielectron} using 
 two templates of open-charm and open-beauty production, while keeping the light-flavor and J$/\psi$ contributions fixed. The results were obtained for the intermediate invariant mass region $1.1<m_{\rm{e^+e^-}}<2.7$~GeV/$c^2$ and for $p_{\rm T} < 8$~GeV/$c$. The differences between the two generators are similar to those from datasets at 7~TeV and 13~TeV.

\section{Multiplicity dependence of charmonium production}
\begin{figure}[h]
	\begin{minipage}{\textwidth}
		\centering
		\raisebox{-0.009\height}{\includegraphics[width=0.397\textwidth]{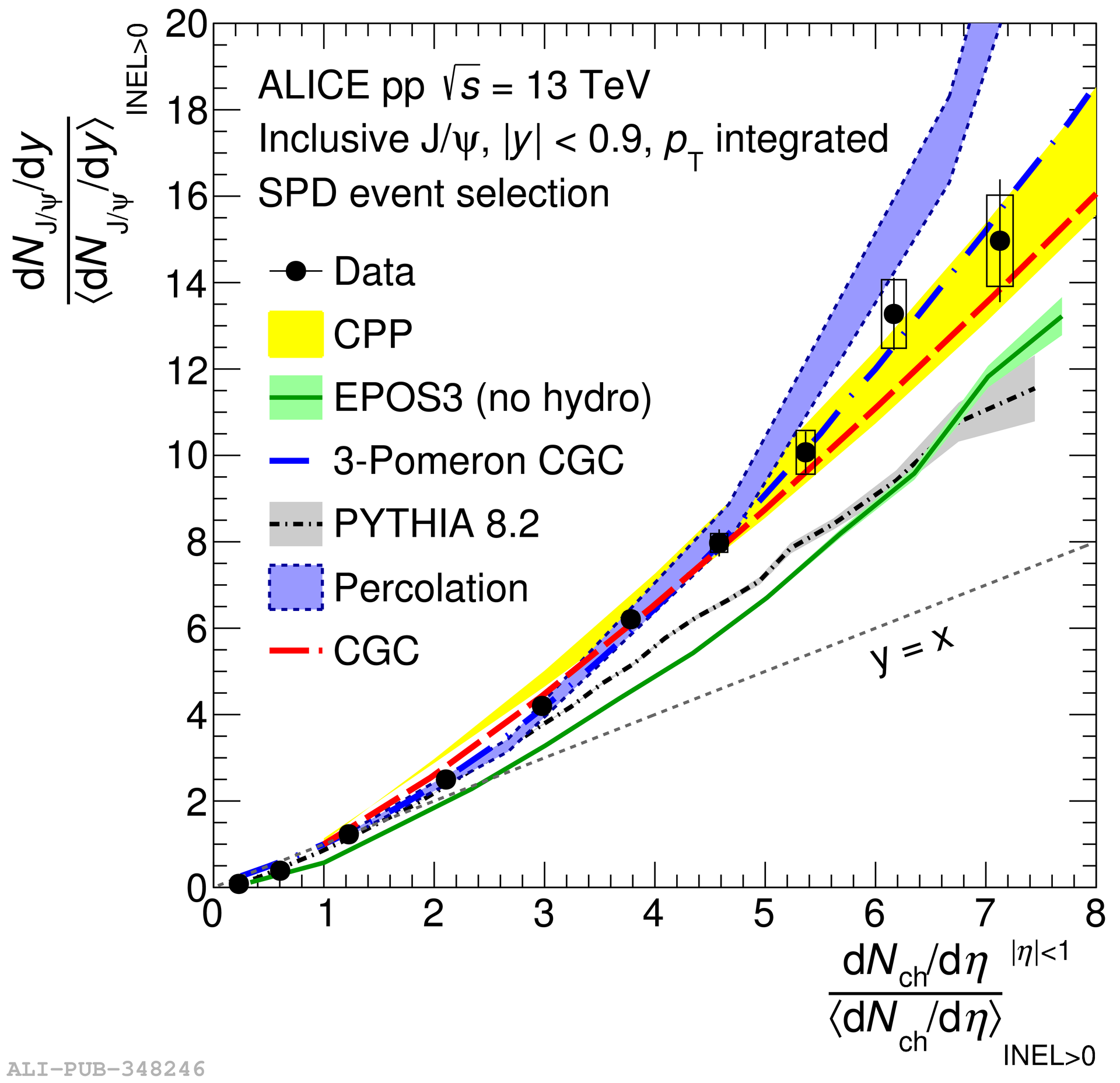}}
		\hspace{0.2cm}
		\raisebox{-0.0\height}{\includegraphics[width=0.49\textwidth]{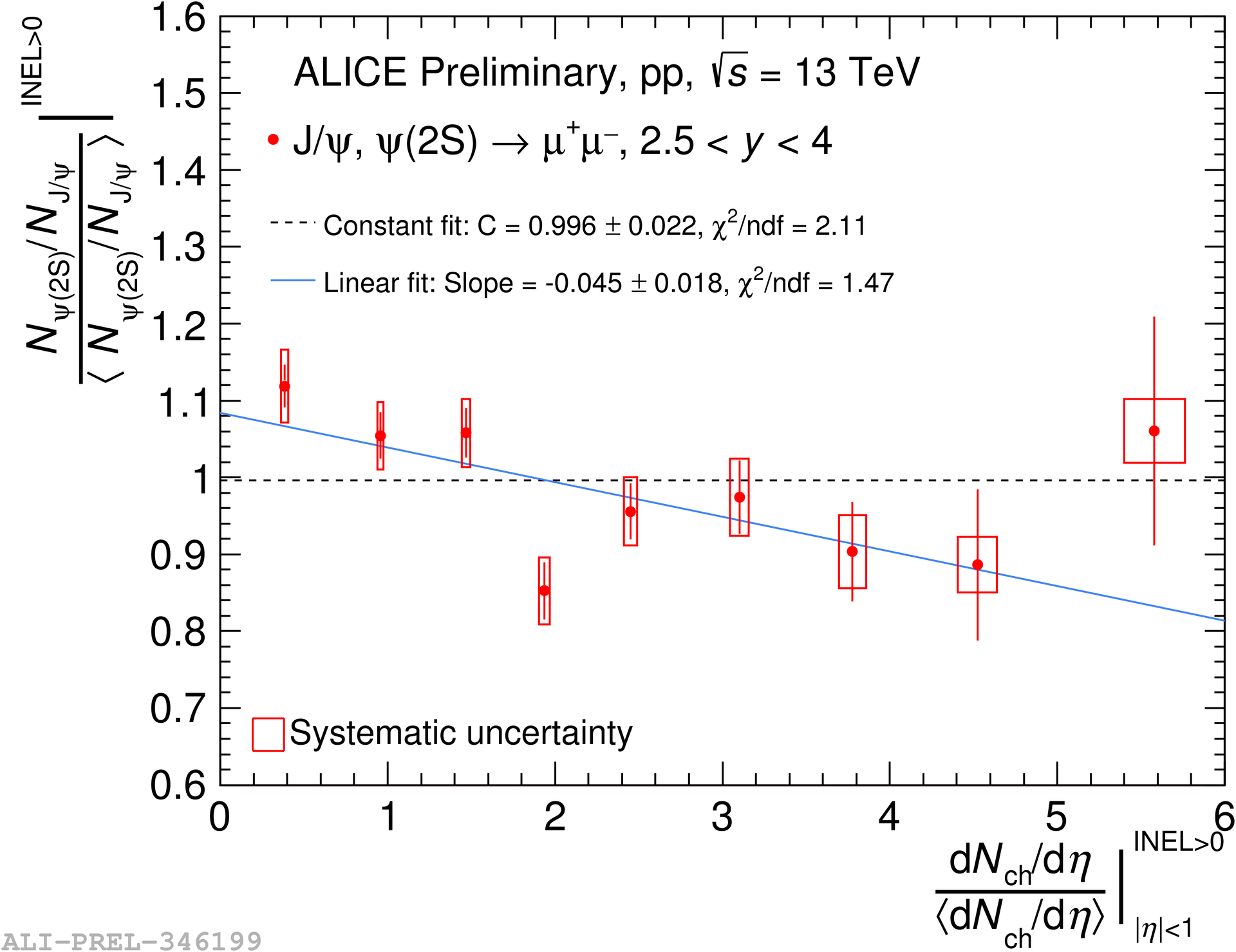}}
	\end{minipage}
	\caption{(Left) The J/$\psi$ self-normalized yield at midrapidity as a function of the self-normalized charged-particle multiplicity density compared to model calculations \cite{Kopeliovich_2013, Werner_2014, Siddikov:2019xvf, Weber_2019,  Ferreiro:2012fb, PhysRevD.98.074025}. (Right) The self-normalized ratio of the $\psi \rm (2S)$ to the J/$\psi$ yield at forward rapidity as a function of the self-normalized charged-particle multiplicity density.}
	\label{fig:Charmonia}
\end{figure}

The interplay between charmonium production and soft processes driving the charged-particle multiplicity is complex and not fully understood. A stronger than linear increase of the J/$\psi$ yield at midrapidity as a function of multiplicity \cite{collaboration2020multiplicity} is qualitatively described by models with different underlying physics mechanisms, as shown in Fig.~\ref{fig:Charmonia} (left). The multiplicity dependence in the presented models arises from a reduction of the multiplicity rather than from an enhancement of the J/$\psi$ production. Three models quantitatively describe the data: Coherent Particle Production (CPP) \cite{Kopeliovich_2013}, Color Glass Condensate (CGC), \cite{PhysRevD.98.074025} and 3-Pomeron CGC \cite{Siddikov:2019xvf}. Nevertheless, none of the models considers non-prompt J/$\psi$ contributions originating from beauty-hadron decays and its association with soft particle production which might play an important role. As implemented in PYTHIA 8.2 
\cite{Weber_2019}, auto-correlation mechanisms such as beauty-quark fragmentation and associated jet production lead to a stronger than linear increase. 

Recently, the $\psi \rm (2S)$ yield was measured as a function of the multiplicity at forward rapidity. Its ratio to the yield of the ground state J/$\psi$ is shown in Fig.~\ref{fig:Charmonia} (right). The measurement suggests that $\psi \rm (2S)$ production increases with the multiplicity more slowly compared to J/$\psi$ although the initial-state models are expected to give rather a similar multiplicity dependence for both states. However, a more precise measurement is needed to draw stronger physics conclusions.

The ALICE apparatus is undergoing major upgrades allowing to collect more data due to increased readout rates (50--100 larger minimum-bias Pb--Pb and 10 times larger high-multiplicity pp data samples than in Run 2). Tracking performance will be improved and secondary vertices will be reconstructed also at forward rapidity. The upgrades will allow heavy-flavor measurements with a higher precision down to $p_{\mathrm{T}} = 0$~GeV/$c$ and a better separation of prompt and non-prompt components.

\end{document}